**Active Control of Bound States in the Continuum in Toroidal Metasurfaces**

*Fedor V. Kovalev\*, Andrey E. Miroshnichenko, Alexey A. Basharin, Hannes Toepfer, and Ilya V. Shadrivov*

F. V. Kovalev, I. V. Shadrivov
ARC Centre of Excellence for Transformative Meta-Optical Systems (TMOS), Research School of Physics, The Australian National University, Canberra ACT 2601, Australia
E-mail: fedor.kovalev@anu.edu.au, ilya.shadrivov@anu.edu.au

A. E. Miroshnichenko
School of Engineering and Technology, University of New South Wales Canberra, Canberra ACT 2612, Australia
E-mail: andrey.miroshnichenko@unsw.edu.au

A. A. Basharin
Department of Physics and Mathematics, Center for Photonics Sciences, University of Eastern Finland, Joensuu 80101, Finland
E-mail: alexey.basharin@uef.fi

H. Toepfer
Advanced Electromagnetics Group, Technische Universität Ilmenau, Ilmenau 98693, Germany
E-mail: hannes.toepfer@tu-ilmenau.de

The remarkable properties of toroidal metasurfaces, featuring ultrahigh-Q bound states in the continuum (BIC) resonances and nonradiating anapole modes, have garnered significant attention. The active manipulation of quasi-BIC resonance characteristics offers substantial potential for advancing tunable metasurfaces. Our study explores explicitly the application of vanadium dioxide, a phase change material widely used in active photonics and room-temperature bolometric detectors, to control quasi-BIC resonances in toroidal metasurfaces. The phase change transition of vanadium dioxide occurs in a narrow temperature range, providing a large variation in material resistivity. Through heating thin film patches of vanadium dioxide integrated into a metasurface comprising gold split-ring resonators on a sapphire substrate, we achieve remarkable control over the amplitude and frequency of quasi-BIC resonances due to their high sensitivity to losses present in the system. Breaking the symmetry of meta-atoms reveals enhanced tunability. The predicted maximum change in the quasi-BIC resonance amplitude reaches 14 dB with a temperature variation of approximately 10 °C.

# 1. Introduction

Over the past fifteen years, significant progress has been made in the study of toroidal metamaterials[1,2]. They were first introduced in 2007[3], with experimental demonstrations of toroidal dipole response occurring in the microwave range in 2010[4]. Subsequent investigations have extended their realization to higher frequencies, including the optical range[5,6].

In metamaterials, the excitation of the toroidal dipole, a component of the multipole expansion[7], is intricately tied to the specific design and arrangement of subwavelength meta-atoms. Special design of the current distribution within meta-atoms is essential to induce the toroidal dipole moment. The required configuration of current loops can be achieved by carefully engineering the meta-atoms' geometry and material properties.

Toroidal metamaterials have found applications in many fields, including detectors, sensors, modulators, emitters, and waveguides[2]. One of their fascinating aspects lies in the ability to excite resonances with remarkably high-quality factors and narrow bandwidth[8–11]. By adjusting electric and toroidal dipole components, one can create an anapole state when their radiation is cancelled in the far field[5,8,12,13]. Metamaterials with anapole configurations can exhibit unique properties, such as near-zero scattering or enhanced field confinement, which is utilized in cloaking devices and advanced optics.

In the realm of metamaterials, bound states in the continuum (BIC)[14–17] refer to specific wave configurations that maintain localization within the material, showcasing zero bandwidth and maintaining complete confinement without any radiation. This becomes possible due to several mechanisms[6,18], for example, due to the symmetry of the mode not allowing the coupling to propagating waves. Recent studies have shown that this phenomenon is one of the ways to achieve very high Q-factors[19,20]. There is a critical distinction between anapoles and non-radiating eigenmodes, such as bound states in the continuum or embedded eigenstates. Unlike bound states in the continuum, anapole is not an eigenmode of the cavity, it cannot sustain oscillations on its own to meet boundary conditions, and can be observed only in scattering[18]. Instead, anapole represents a resonant distribution of fields that can be activated by a specific impinging excitation.

The concept of the toroidal dipole BIC was initially introduced in all-dielectric optical metasurfaces[21]. Symmetry breaking in silicon metasurfaces was demonstrated to excite ultrahigh-Q toroidal dipole leaky resonances. More recently, it was shown that high Q-factors of metal toroidal metamaterials can also be explained by using the concept of BIC[9,22]. Toroidal dipole BIC metasurfaces stand out as promising candidates for ultrasensitive measurements due to their substantial changes in resonance frequency and amplitude with the slightest variation in the surrounding media. They provide a platform for a strong electric and magnetic field enhancement and can be used for creating compact lasers.

Active control of toroidal dipole quasi-BIC resonances represents a promising avenue for further exploration and research. Recently, dynamic tuning of toroidal metasurface characteristics has been achieved using thin films of silicon[23], graphene[24–27], phase-change materials such as GST (Germanium-Antimony-Tellurium)[28] and $VO_2$ (vanadium dioxide)[29–31], and other techniques, for example, integration of diodes[32] or utilization of microelectromechanical systems[33,34]. Tunable toroidal metamaterials have unlocked new opportunities for resonant applications and multifunctional devices with remarkable potential[2].

Recently, a concept of metabolometers has been introduced, leveraging the unique properties of toroidal metamaterials[35]. The metabolometer combines a high-Q factor toroidal metamaterial with an embedded micro-pad superconductor, serving as an absorber for terahertz radiation. The practical implementation involves a terahertz beam heating the absorber while a plane gigahertz wave wirelessly reads out the absorber's temperature, unveiling a novel approach with potential implications for advancing terahertz/millimeter-wave detectors in the future.

$VO_2$ stands out as a promising material in the realms of both bolometric detection[36–43] and active photonics[44–46] thanks to its distinctive semiconductor-to-metal transition (SMT) near room temperature. Bolometers are widely utilized devices for measuring incident electromagnetic radiation energy[37]. They rely on changes in the electrical resistance of a thermosensitive absorbing material when it is heated due to the absorption of an incident wave. $VO_2$'s sharp transition and reversible phase change enable sensitive temperature-dependent measurements, facilitating rapid and accurate detection of incident radiation. The material's broad sensitivity across wavelengths makes it valuable for applications like infrared and terahertz imaging. In active photonics, $VO_2$'s tunable optical properties and switchable characteristics offer dynamic control over refractive indices and the modulation of light transmission or reflection.

$VO_2$ is a remarkable material known for its intriguing phase change transition. At temperatures below approximately 68 degrees Celsius, $VO_2$ is in its semiconducting phase, characterized by a distorted monoclinic crystal structure. As the temperature rises beyond this point, a rapid and reversible phase transition occurs, leading to the metallic phase due to a tetragonal crystal structure. This transition is accompanied by significant changes in electrical and optical properties, making VO$_2$ a valuable material for various applications. A crucial parameter in comprehending VO$_2$'s behavior during the phase transition is the large temperature coefficient of resistance (TCR). The substantial TCR is pivotal for applications requiring sensitive temperature-dependent measurements, and underscores VO$_2$'s suitability for tunable devices.

In this study, we propose that thin film patches of VO$_2$ integrated into the toroidal metasurface may dramatically change the amplitude and frequency of quasi-BIC resonances as the temperature changes. The proposed tunable metasurface comprises an array of meta-atoms formed by two symmetrical gold split-ring resonators on a sapphire substrate. The phase change transition of VO$_2$ occurs in a narrow temperature range, providing a large variation in its conductivity. This leads to a change in the transmission at the frequency of the quasi-BIC resonance and, accordingly, in the amplitudes of the multipole expansion components, demonstrating a high sensitivity to the losses present in the system. Additionally, we analyse the behavior of quasi-BIC resonances when introducing asymmetry in the geometry of the meta-atom. The amplitude of resonances increases with the degree of asymmetry, corresponding to the excitation of symmetry-protected quasi-BIC resonances. The proposed tunable metasurface holds potential applications in various fields, including active photonic systems, room-temperature terahertz and infrared bolometers, as well as multifunctional metadevices.

## 2. Bound states in the continuum in toroidal metasurfaces

We study the metasurface in the form of a square array of meta-atoms illustrated in Figure 1a. Meta-atoms are shown in Figure 1b, and in this study we manipulate their central and outer gaps. The size of the unit cell $p$ is 475 μm x 475 μm with the central gap size $dc$ = 40 μm. The outer radius of the resonator $R$ = 235 μm, the track width $w$ = 30 μm, the thickness of the resonator $t$ = 0.51 μm. The outer gap size $dl$ is used to tune the quasi-BIC resonance. The linearly polarized wave with the E-field parallel to the central microstrip is normally incident on the metasurface.

We first study the symmetrical meta-atoms without a substrate. We use the Drude model for gold in the terahertz range[47]. Figure 1c shows the quasi-BIC resonance for different outer gaps of the resonators. The proposed metasurfaces can support quasi-BIC resonances by tuning structure parameters. When $dl$ = 50 μm, the resonance disappears, corresponding to the accidental BIC, which has zero bandwidth and infinite Q-factor in the ideal case of perfect electric conductors. We analyzed the vanishing resonance at 314.8 GHz for *dl* = 50 μm using the Eigenmode solver in CST Studio. Surface currents of the BIC eigenmode in the meta-atom are shown in Figure S1 (Supporting Information).

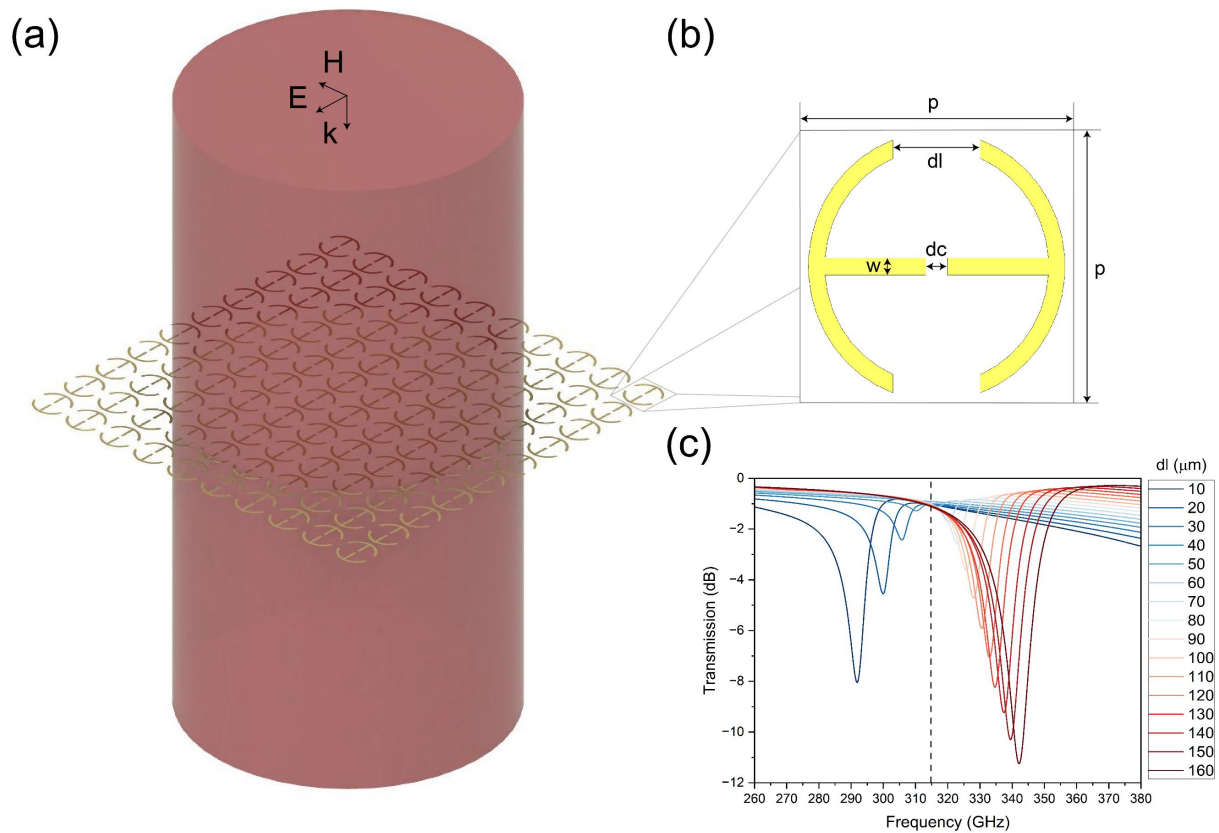

**Figure 1.** Schematics of the problem with the EM wave incident on the metasurface (a); geometry of the meta-atom (b); transmission coefficient for different outer gap sizes (c). The dotted vertical line indicates the BIC at approximately 315 GHz.

Any losses dramatically reduce the amplitude of such quasi-BIC resonances that can have an extremely high Q-factor ($\geq 10^4$) in ideal metasurfaces composed of perfect electric conductors. We observe that losses in gold reduce the quality factor, so the resonance band becomes much wider, and its amplitude decreases. For example, the transmission of the proposed metasurface at the quasi-BIC resonance frequency is around -6 dB in the case of gold and below -50 dB in the case of perfect electric conductors for the same size of the outer gap *dl* = 100 μm.

Figure 2a shows that the electric field at the resonant frequency (242 GHz) for the outer gap size of 160 μm is mostly localized in the central gap of the meta-atoms. Figure 2b shows the magnetic field at the resonant frequency. Figure 2c shows currents at the resonant frequency creating two loops with opposite flow directions. Such configuration of currents, as well as electric and magnetic fields corresponds to the excitation of the toroidal dipole[8].

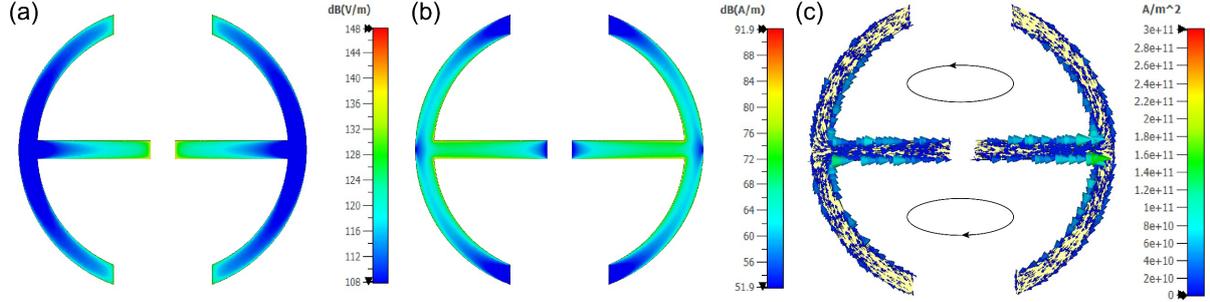

**Figure 2.** Electric (a) and magnetic (b) fields and current density (c) within the designed meta-atom at the quasi-BIC resonance frequency for the outer gap size (*dl*) equal to 160 μm.

Figure 3a and 3b shows reflection and transmission of the metasurface for *dl* = 50 μm and 160 μm. Power scattered by individual multipoles shown in Figures 3c and 3d was calculated by integrating the current density over the volume using the following formulas for the exact multipole moments[48–50]:

$$p_\alpha = -\frac{1}{i\omega} \int J_\alpha j_0(kr) d^3r, \tag{1}$$

$$m_\alpha = \frac{3}{2} \int (\mathbf{r} \times \mathbf{J})_\alpha \frac{j_1(kr)}{kr} d^3r, \tag{2}$$

$$T_\alpha^{(e)} = -\frac{k^2}{2i\omega} \int [3(\mathbf{r} \cdot \mathbf{J}) r_\alpha - r^2 J_\alpha] \frac{j_2(kr)}{(kr)^2} d^3r, \tag{3}$$

$$Q_{\alpha\beta}^{(e)} = -\frac{3}{i\omega} \int [3(r_\alpha J_\beta + r_\beta J_\alpha) - 2\delta_{\alpha\beta}(\mathbf{r} \cdot \mathbf{J})] \frac{j_1(kr)}{kr} d^3r, \tag{4}$$

$$T_{\alpha\beta}^{(Qe)} = -\frac{6k^2}{i\omega} \int [5 r_\alpha r_\beta (\mathbf{r} \cdot \mathbf{J}) - (r_\alpha J_\beta + r_\beta J_\alpha) r^2 - r^2 (\mathbf{r} \cdot \mathbf{J}) \delta_{\alpha\beta}] \frac{j_3(kr)}{(kr)^3} d^3r, \tag{5}$$

$$Q_{\alpha\beta}^{(m)} = 15 \int [(\mathbf{r} \times \mathbf{J})_\alpha r_\beta + (\mathbf{r} \times \mathbf{J})_\beta r_\alpha] \frac{j_2(kr)}{(kr)^2} d^3r, \tag{6}$$

$$O_{\alpha\beta\gamma}^{(e)} = \frac{15i}{\omega} \int (r_\gamma r_\beta J_\alpha + r_\gamma r_\alpha J_\beta + r_\alpha r_\beta J_\gamma - A_{\alpha\beta\gamma}) \frac{j_2(kr)}{(kr)^2} d^3r, \tag{7}$$

$$P_{scat} = \frac{k^4}{12\pi\varepsilon_0^2 c\mu_0} |p + T^{(e)}|^2 + \frac{k^4}{12\pi\varepsilon_0 c} |m|^2 + \frac{k^6}{1440\pi\varepsilon_0^2 c\mu_0} |Q^{(e)} + T^{(Qe)}|^2 +$$

$$+ \frac{k^6}{1440\pi\varepsilon_0 c} |Q^{(m)}|^2 + \frac{k^8}{3780\pi\varepsilon_0^2 c\mu_0} |O^{(e)}|^2, \tag{8}$$

where $p_\alpha$ is the electric dipole moment, $m_\alpha$ is the magnetic dipole moment, $T_\alpha^{(e)}$ is the toroidal electric dipole moment, $Q_{\alpha\beta}^{(e)}$ is the electric quadrupole moment, $T_{\alpha\beta}^{(Qe)}$ is the toroidal electric quadrupole moment, $Q_{\alpha\beta}^{(m)}$ is the magnetic quadrupole moment, $O_{\alpha\beta\gamma}^{(e)}$ is the electric octupole moment, $A_{\alpha\beta\gamma} = \delta_{\alpha\beta}V_\gamma + \delta_{\beta\gamma}V_\alpha + \delta_{\alpha\gamma}V_\beta$, $\mathbf{V} = \frac{1}{5}[2(\mathbf{r}\cdot\mathbf{J})r_\alpha - r^2 J_\alpha]$, $j_n(kr)$ are the spherical Bessel functions of the first kind of the order $n$, $\omega$ is the frequency, $\mathbf{J}$ is the current density, $\mathbf{r}$ is the radius vector, $\delta$ is the Kronecker delta symbol, $k$ is the wave number, $c$ is the speed of light, $\varepsilon_0$ and $\mu_0$ are the electric and magnetic constants, indices $\alpha$, $\beta$, $\gamma$ correspond to $x$, $y$, $z$ components. In this work, the center of coordinates for the purpose of the multipole expansion is chosen at the middle of the unit cell.

To verify the accuracy of the multipole expansion, Figures 3a and 3b also show the reflection and transmission of the metasurface calculated from the exact multipole moments using Eq. 1-8, S1 and S2. Good agreement with the results of full numerical simulations demonstrates the high accuracy of the multipole decomposition.

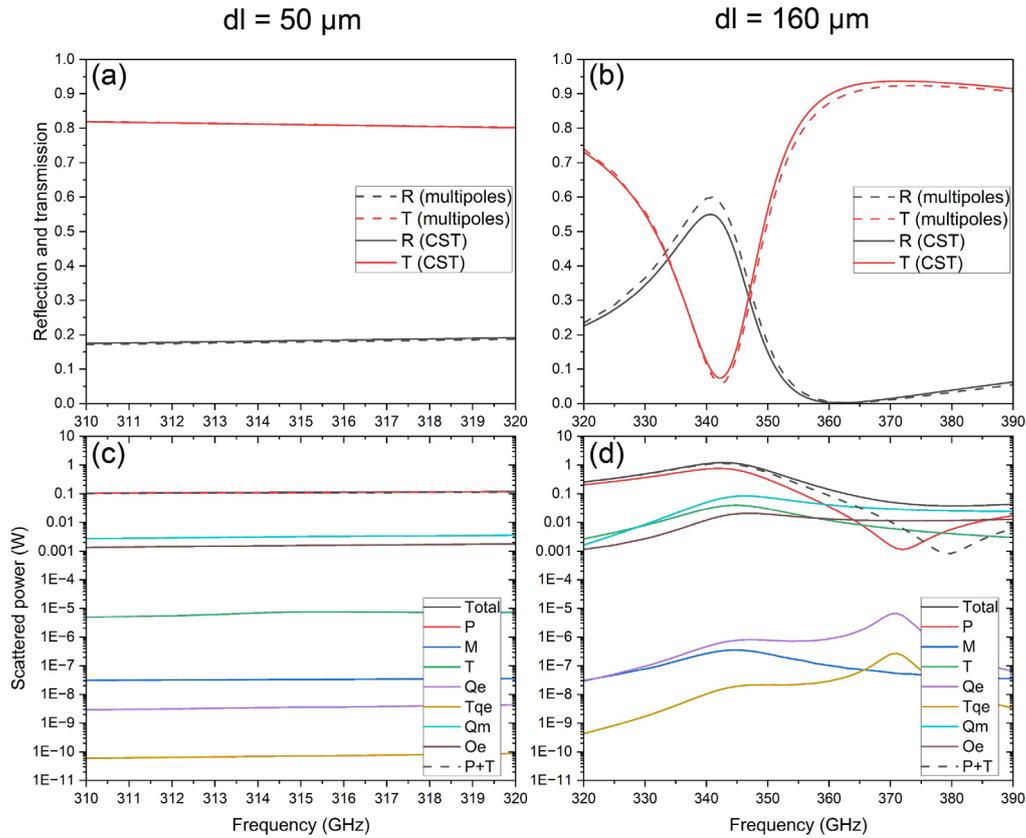

**Figure 3.** Reflection and transmission of the metasurface obtained in CST Studio (solid) and calculated from the multipole decomposition (dashed) and power scattered by multipoles for the outer gap size $dl = 50$ μm (a, c) and $dl = 160$ μm (b, d).

We observe a four orders of magnitude increase in the scattered power by the toroidal dipole (T) near the quasi-BIC resonance in Figures 3c and 3d when changing the outer gap value *dl* from 50 to 160 µm, respectively. The power scattered by the electric dipole (P) remains significantly higher than the power scattered by the toroidal dipole (T) and the magnetic quadrupole (Qm). The magnetic quadrupole (Qm) plays a parasitic role in such toroidal metamaterials[8]. We also found that the electric octupole (Oe), which has not been considered in previous studies, plays an important role in the excitation of the quasi-BIC resonances. The power scattered by other multipoles (the magnetic dipole M, the electric quadrupole Qe and the toroidal electric quadrupole Tqe) is significantly lower. The obtained results agree with previous studies of toroidal metamaterials[8,9].

Comparing the reflection and transmission characteristics (directional) with the scattered power by multipoles (non-directional) does not allow for a clear conclusion about their contribution to the metasurface characteristics. However, the transmission minimum frequency in Figure 3b corresponds to the maximum overall scattered power by multipoles in Figure 3d. We provide the reconstruction of the reflection and transmission characteristics from the multipole components[51] in Figure 3a,b and compare their contribution in Figure S2 (Supporting Information).

The toroidal dipole component plays a substantial role in the excitation of the quasi-BIC resonance. It should be noted that the quasi-BIC resonances and the metasurface in this study can be referred to as toroidal due to the possibility of making the toroidal component dominant (the same order as the electric dipole) by carefully optimizing the meta-atom structure and using metals with low losses[8,9]. This would also lead to a very high Q-factor and shift the frequency of the anapole state (the destructive interference of the electric and toroidal dipoles: P+T minimum in Figure 3d) close to the BIC shown in Figure 1. However, achieving this is beyond the scope of this article.

We analyzed the quality factor of the excited quasi-BIC resonance by fitting the transmission spectrum with a Fano formula[52]:

$$T_{Fano}(\omega) = t_{d_0} + t_d^2 \frac{\left[q + \frac{2(\omega - \omega_0)}{d\omega}\right]^2}{1 + \frac{4(\omega - \omega_0)^2}{d\omega^2}}, \qquad (9)$$

where $t_{d_0}$ is the minimum of transmission near resonance, $t_d$ is the non-resonant transmission amplitude, $q$ is the shape factor, $\omega_0$ is the resonance frequency, $d\omega$ is the resonance bandwidth.

The Q-factor can be found as $Q = \frac{\omega_0}{d\omega}$ and is found to be 27.5 when $dl = 160$ μm. The dependence of the Q-factor and bandwidth of the quasi-BIC resonance on the outer gap size $dl$ and the collision frequency is detailed in Figure S3 (Supporting Information). Notably, the Q-factor exceeding $10^3$ for a toroidal dipole quasi-BIC resonance in a metasurface composed of aluminium split-ring resonators on a 50 μm-thick cyclic olefin copolymer substrate was recently reported in the terahertz range[9].

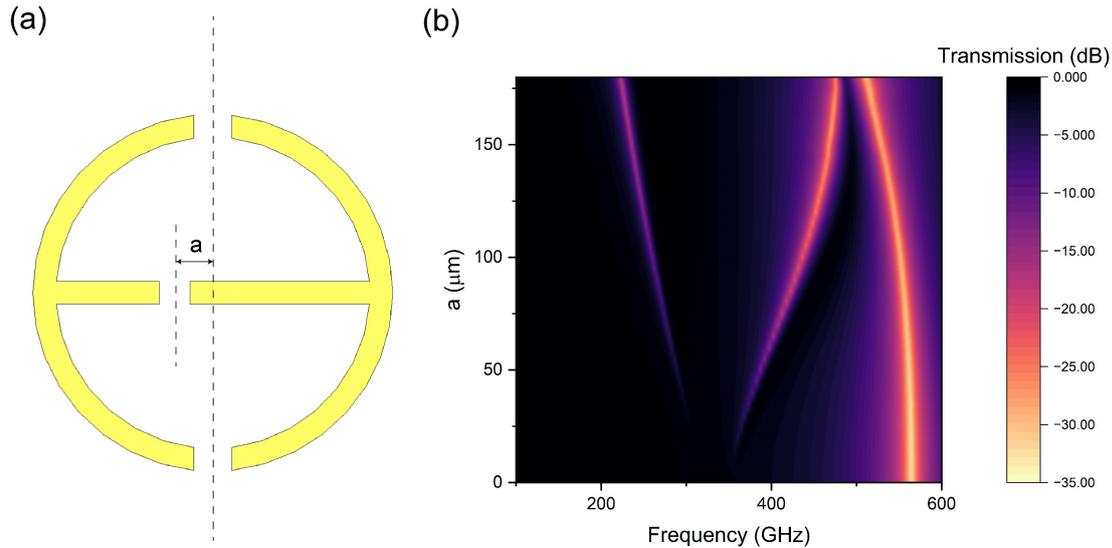

**Figure 4.** Asymmetric meta-atom with an offset parameter $a$ (a) and the metasurface transmission spectrum for different $a$ (b).

Next, we study the behavior of quasi-BIC resonances when the central gap is shifted from the center of the meta-atom, as shown in Figure 4a. The transmission characteristics for the different offset $a$ for the fixed outer gap size $dl = 50$ μm are shown in Figure 4b. When $a = 0$ μm, there is no resonance. We found that two resonances appear in the asymmetric structure when $a > 0$ μm. Asymmetrical meta-atoms in gold metasurfaces can induce quasi-BIC resonances through symmetry breaking, and their amplitude increases with the asymmetry parameter.

Figures 5a and 5b show the transmission and reflection characteristics of the metasurface with broken symmetry obtained in CST Studio and calculated from the exact multipole moments using Eq. 1-8, S1 and S2 for two values of the asymmetry parameter: 50 μm and 180 μm. Power scattered by multipoles is shown in Figures 5c and 5d. Both resonances − the first at 289 GHz ($a = 50$ μm) and 223 GHz ($a = 180$ μm) and the second at 387 GHz ($a = 50$ μm) and 475 GHz ($a = 180$ μm) − notably feature a robust contribution from the electric quadrupole ($Q_e$).

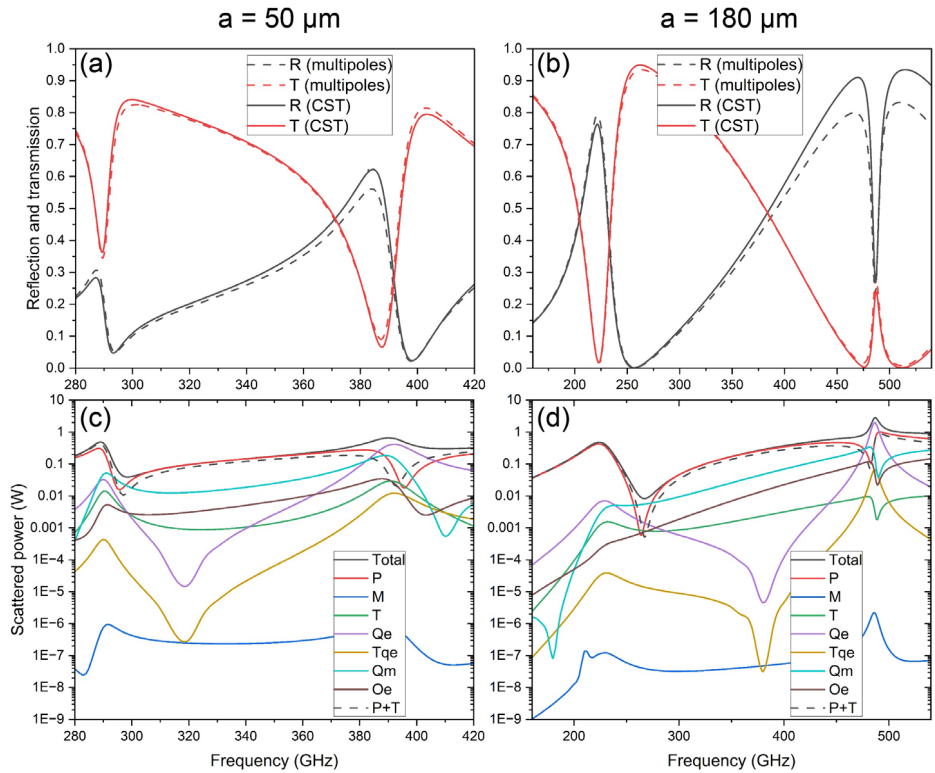

**Figure 5.** Transmission and reflection characteristics of the metasurface with broken symmetry obtained in CST Studio (solid) and calculated from the multipole decomposition (dashed) and power scattered by multipoles for two offset parameters $a$: 50 μm (a,c) and 180 μm (b,d).

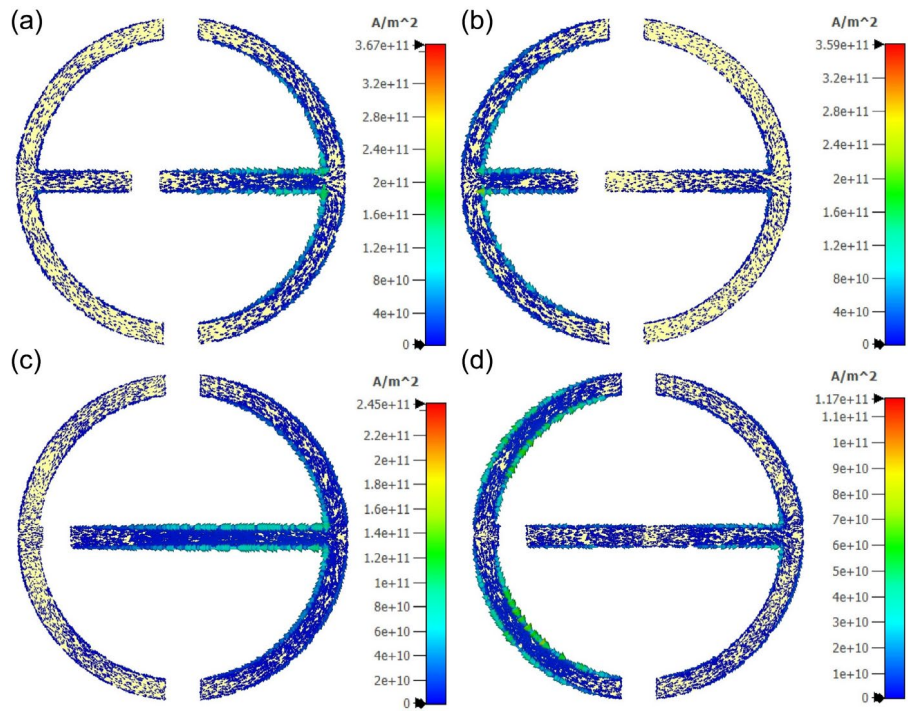

**Figure 6.** Current density in the meta-atom with broken symmetry for two offset parameters $a$: 50 μm at 289 GHz (a) and 387 GHz (b), 180 μm at 223 GHz (c) and 475 GHz (d) respectively.

Figure 6 illustrates the current density in the meta-atom with broken symmetry for two parameters of asymmetry *a*: 50 μm at 289 GHz (a) and 387 GHz (b); 180 μm at 223 GHz (c) and 475 GHz (d), respectively. The currents at the lower resonant frequency have a maximum in the right section, while those at the higher resonant frequency have a maximum in the left section. This behavior can be explained by the difference in the length of the central microstrip when the symmetry is broken.

## 3. Active toroidal metasurfaces and control of bound states in the continuum

Next, we study a metasurface consisting of gold split-ring resonators with integrated thin film patches of VO$_2$. The Drude model can be employed to characterize the permittivity of VO$_2$ within the terahertz range[29,53], providing an effective representation of charge carriers and aiding researchers in predicting dynamic changes in resistivity and conductivity associated with the phase transition:

$$\varepsilon(\omega) = \varepsilon_\infty - \frac{\omega_p^2(\sigma)}{\omega^2 + i\gamma\omega}, \tag{10}$$

where $\varepsilon_\infty$ is the permittivity at high frequencies, $\omega_p(\sigma)$ is the conductivity dependent plasma frequency and $\gamma$ is the collision frequency. The plasma frequency can be approximately expressed as $\omega_p^2(\sigma) = \frac{\sigma}{\sigma_0}\omega_p^2(\sigma_0)$. We used the following constants in our modeling[53]: $\sigma_0 = 3 \times 10^5$ S/m, $\varepsilon_\infty = 12$, $\omega_p(\sigma_0) = 1.4 \times 10^{15}$ rad/s, $\gamma = 5.75 \times 10^{13}$ rad/s.

For the final metasurface design we used the substrate with the following characteristics: $\varepsilon_r = 11.5$ and tan δ = 0.0001, which corresponds to sapphire. The thickness of the substrate was chosen to be 1 μm to avoid Fabry-Perot resonances, as well as to obtain a higher Q-factor. We chose the outer gap size of 160 μm for the integration of VO$_2$ thin film patches due to the significant quasi-BIC resonance amplitude (over 10 dB). The central gap can be loaded with a tuning element, a thin film patch (100 nm thick) of VO$_2$ sized at 40 μm x 5 μm. The tunable meta-atom is shown in Figure 7a. Modulating the electrical conductivity of the film by varying the material's temperature induces changes in the transmission coefficient and phase of transmitted waves, as illustrated in Figures 7c and 7e. Our study was guided by the following experimental data obtained for 107-nm-thick VO2 films on sapphire[41]: the conductivity values for temperatures of 70°C (close to the maximum temperature coefficient of resistance) and 80°C were measured as 1000 S/m and 100000 S/m, respectively.

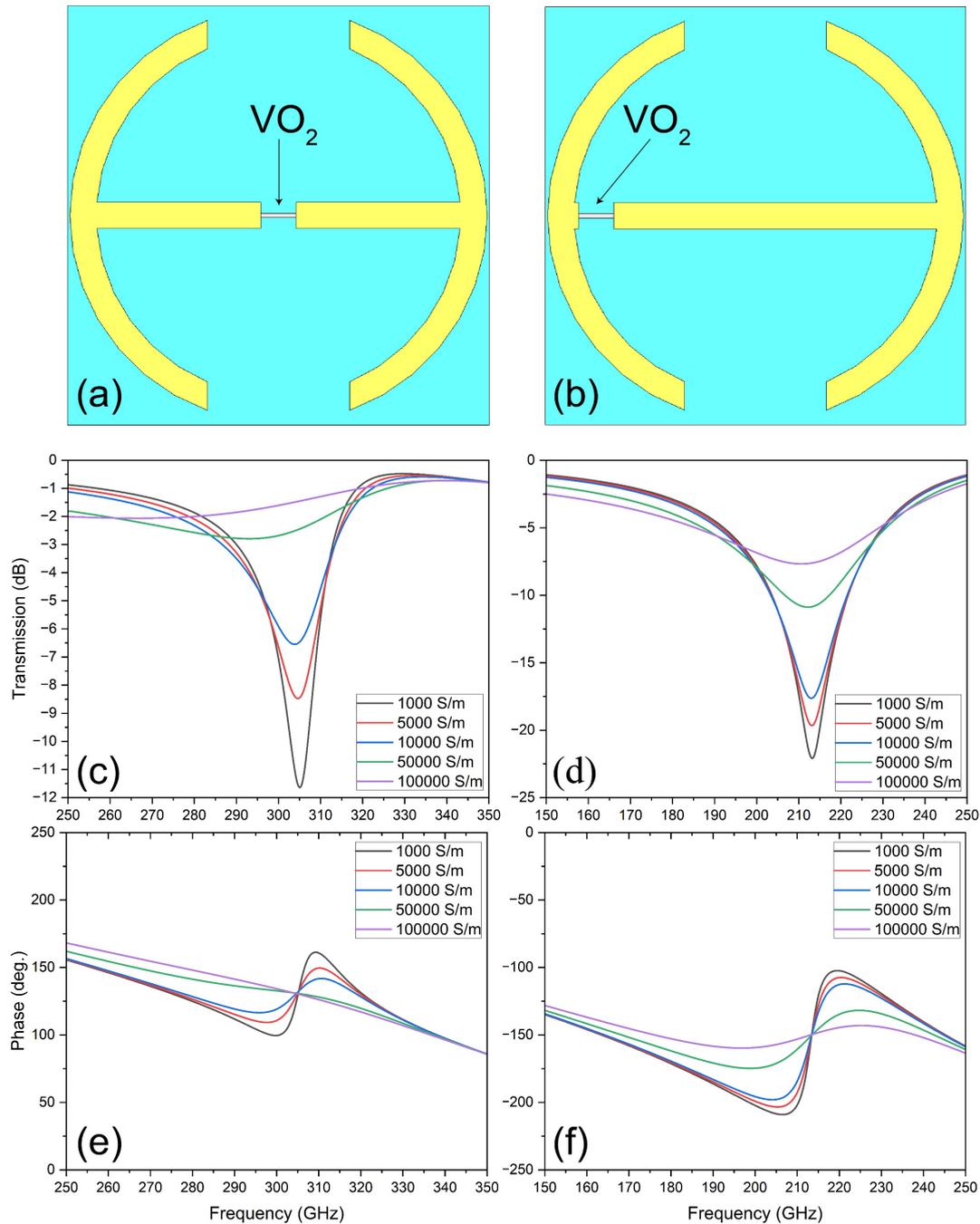

**Figure 7.** Tunable symmetric (a) and asymmetric (b) meta-atoms with integrated thin film patches of $VO_2$ in the central gap and simulated transmission characteristics (c,d) of the corresponding metasurfaces and phase of the transmitted waves (e,f) for different values of $VO_2$ conductivity.

Heating by the incident terahertz or infrared radiation can switch VO$_2$ films between two phases – semiconducting and metallic. This leads to a change in the electric and magnetic field at the probing frequency of the quasi-BIC resonance and, accordingly, in the amplitudes of the multipole expansion components when a subterahertz electromagnetic wave is incident on the developed toroidal metasurface. For this reason, the quasi-BIC resonance changes its amplitude, demonstrating a high sensitivity to the losses.

Symmetry breaking can increase the quasi-BIC resonance amplitude. The asymmetric meta-atom of the tunable toroidal metasurface with the central gap shift *a* equal to 180 μm is shown in Figure 7b. The asymmetry allows to increase the transmission change of the quasi-BIC resonance by more than 3 dB. The transmission coefficient and phase of transmitted waves for different values of electrical conductivity of the integrated VO$_2$ film is shown in Figs. 7d and 7f, respectively. Figure S4 (Supporting Information) illustrates the current density within the symmetric and asymmetric meta-atoms with the integrated VO$_2$ patches for their different conductivity values.

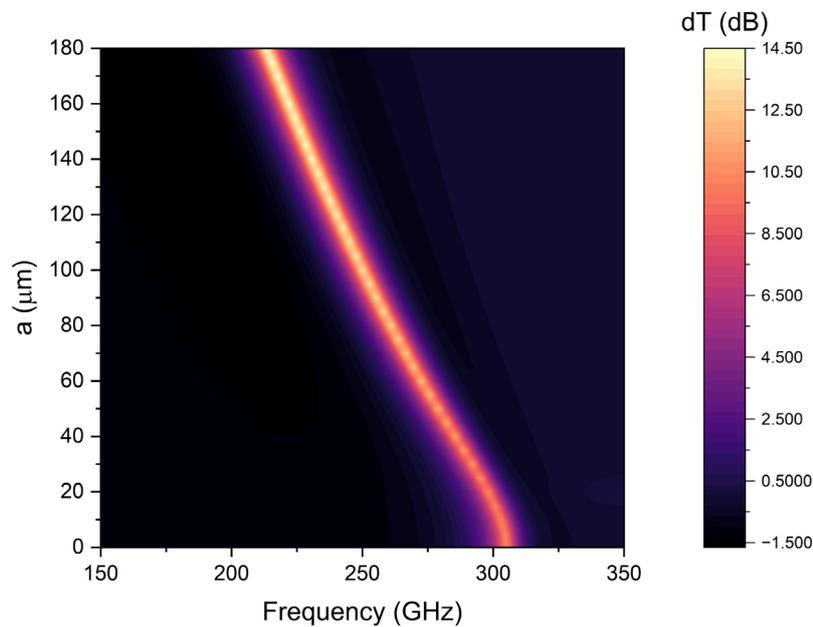

**Figure 8.** Transmission change for different asymmetry parameters.

We conclude that it is possible to use the asymmetry to control the quasi-BIC resonance behavior and improve the metasurface tunability. Change in the transmission spectrum for different asymmetry parameters is shown in Fig. 8. It is calculated as the difference between the transmission coefficients when VO$_2$ switches from the semiconducting to the metallic states. The difference reaches its maximum when the central gap is maximally shifted from its symmetrical position. The maximum change of the transmission coefficient is predicted to be 14 dB as the temperature changes by approximately 10 °C.

We conduct an analysis of the active toroidal metasurface characteristics in scenarios where the integrated thin film patches of $VO_2$ lack a galvanic connection to the split-ring resonators. In this configuration, the quasi-BIC resonance exhibits both frequency and amplitude shifts, shown in Figure S5 (Supporting Information), albeit with a smaller amplitude change at a given frequency compared to the previously discussed case. This design, despite the reduced amplitude change, holds promise for bolometric detection, especially given the imperative for thermal isolation of the absorber from other bolometer elements.

## 4. Discussion

Our study explores the intricate behavior of toroidal metasurfaces, focusing on the effects of symmetry breaking and the integration of $VO_2$ thin film patches. The presented modeling results shed light on the principles governing quasi-BIC resonances, the impact of losses on their characteristics, and the tunability of toroidal metasurfaces.

The quasi-BIC resonances exhibited by the proposed metasurface are highly dependent on the geometry of the meta-atoms, specifically the outer and central gaps of the split-ring resonators. Losses in gold are identified as a critical factor affecting the Q-factor of the quasi-BIC resonances, resulting in a wider bandwidth and reduced amplitude. Our observations confirm that losses play a significant role in the performance of toroidal metamaterials, emphasizing the importance of careful meta-atom optimization and the use of low-loss metals or superconductors for achieving ultrahigh-Q factors and the predominance of the toroidal dipole. Our work introduces the asymmetry in the toroidal metasurface design, revealing a fascinating connection to BIC. The amplitude of quasi-BIC resonances increases with the degree of asymmetry, presenting an avenue for manipulating resonance behavior. This insight into symmetry breaking opens possibilities for tailored applications in active photonic systems and bolometric detection. Our work aligns with recent toroidal dipole BIC metasurface studies, highlighting the potential for ultrasensitive measurements and applications in various fields.

The integration of $VO_2$ thin film patches introduces a dynamic tunability to the toroidal metasurface, driven by the semiconductor-to-metal transition of $VO_2$. Our study emphasizes the role of symmetry breaking in enhancing the tunability of quasi-BIC resonances. By introducing asymmetry in the central gap of the meta-atom, we show a substantial increase in transmission change, providing a means to control and improve metasurface tunability. The presented metasurface design, incorporating VO$_2$, demonstrates a significant change in quasi-BIC resonance amplitude with temperature variations, opening possibilities for applications in terahertz and infrared room-temperature bolometers and active photonics.

In contrast to earlier study[29], the quasi-BIC resonance is excited when VO$_2$ is in the semiconducting state. This allows the use of one absorber for the control of the resonance amplitude and frequency. This distinctive feature sets our metasurface design apart, making it particularly suitable for bolometric applications.

The integration of VO$_2$ thin film patches into the resonators allows the use of its dynamic and reversible semiconductor-to-metal transition (SMT) near room temperature. This transition, accompanied by significant changes in electrical and optical properties, enables our metasurface to function effectively as a bolometric readout for the detection of terahertz or infrared radiation, as well as a tunable filter or modulator in future communication systems in the subterahertz range. By attaching antennas to the VO$_2$ film patches, it is possible to enhance the efficiency of terahertz or infrared radiation absorption. Alternatively, this can be achieved by using lenses placed on the absorbers for focusing the radiation onto them. This approach makes it possible to protect VO$_2$ film patches from oxidation.

The concept of metabolometers, as previously developed[35], faced limitations due to the requirement for cryogenic cooling when employing superconductors as absorbing materials. The proposed use of VO$_2$ instead of superconductors eliminates the need for cryogenic conditions and opens the possibility of designing room-temperature metabolometers. This advancement is crucial for practical applications in various scientific and technological domains where the demand for efficient room-temperature detectors is high. Another significant improvement involves adjusting the size of meta-atoms. Specifically, we opted for the subterahertz range to excite the quasi-BIC resonance, departing from the previously suggested microwave readout. This choice aims to enlarge the effective absorption area for incident radiation, as smaller resonator sizes are achieved by elevating the frequency of the quasi-BIC resonance. Our design also highlights the potential for using thin and low-loss substrates for the further experimental implementation of this device.

In essence, this work not only contributes to the understanding of toroidal metamaterials but also offers a practical solution for advancing bolometric applications. The room-temperature capabilities of the designed metasurface pave the way for the development of metabolometers that are not only highly sensitive but also operationally feasible in a broader range of environments. This shift from cryogenic to room-temperature operation represents a significant step forward in the utilization of the proposed toroidal metasurface for real-world applications, particularly in the field of terahertz detection and imaging.

## 5. Conclusion

In conclusion, control of quasi-BIC resonances in toroidal metasurfaces, specifically by incorporating thin-film patches made of vanadium dioxide, reveals a promising avenue for achieving tunable and highly sensitive responses. By tuning structure parameters, breaking symmetry in meta-atoms, and leveraging the temperature-dependent properties of vanadium dioxide, we demonstrated the ability to control the amplitude of high-Q quasi-BIC resonances dynamically. Breaking the symmetry of meta-atoms further expands the tunability range, offering a versatile platform for tailoring quasi-BIC resonances.

The proposed tunable toroidal metasurface exhibits great promise for applications in active photonic systems, room-temperature bolometers, and multifunctional metadevices. The ability to manipulate high-Q quasi-BIC resonances opens doors to innovations in fields ranging from communication to sensing technologies.

## 6. Methods

*Electromagnetic simulations*: the proposed metasurface was modeled as an array of meta-atoms using the frequency domain solver and the Floquet boundary condition in CST Studio. Transmission and reflection of the toroidal metasurface, phase of transmitted waves, electric and magnetic fields, and currents in the meta-atom were calculated using S-Parameters and appropriate monitors within the frequency domain solver. The Q-factor of the excited quasi-BIC resonance was calculated by fitting the transmission spectrum exported from CST Studio with a Fano formula. The BIC eigenmode surface currents in the meta-atom in Supporting Information were found using the Eigenmode Solver in CST Studio.

*Multipole expansion:* power scattered by multipoles was calculated in Matlab using the current density exported from CST Studio.

**Supporting Information**
Supporting Information is available from the Wiley Online Library or from the authors.

**Acknowledgements**

The authors thank Andrey Evlyukhin, Daria Smirnova and Kirill Koshelev for useful discussions. Fedor Kovalev and Ilya Shadrivov were supported by the Australian Research Council Centre of Excellence for Transformative Meta-Optical Systems (Project ID CE200100010). Alexey Basharin was supported by the Academy of Finland via Flagship Programme Photonics Research and Innovation (PREIN), Decision No. 320166, and Grant No. 343393, the Horizon 2020 RISE DiSeTCom Project via No. 823728, the Horizon 2020 RISE CHARTIST Project via No. 101007896, the Horizon 2020 RISE TERASSE Project via No. 823878.

# Supporting Information

**Active Control of Bound States in the Continuum in Toroidal Metasurfaces**

*Fedor V. Kovalev\*, Andrey E. Miroshnichenko, Alexey A. Basharin, Hannes Toepfer, and Ilya V. Shadrivov*

The surface current of the BIC eigenmode at 314.8 GHz within the meta-atom is shown in Figure S1. These results were obtained using the Eigenmode solver in CST Studio. This configuration of toroidal resonators (modeled as PEC) has the following parameters: the outer gap size ($dl$) is 50 µm, and the central gap size ($dc$) is 40 µm. Figure S1 illustrates two current loops with opposing flow directions.

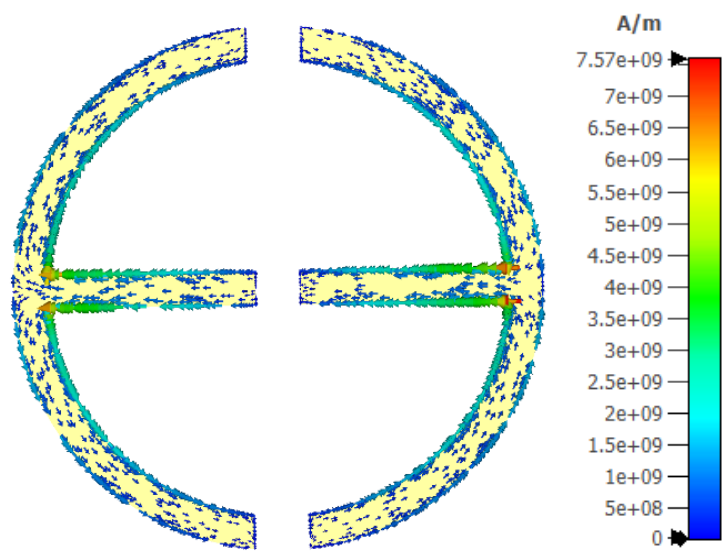

**Figure S1.** Surface currents of the BIC eigenmode in the meta-atom.

Figure S2 shows the reflection and transmission characteristics of the metasurface with the outer gap size $dl$ = 160 µm obtained directly from numerical modeling in CST and calculated from the exact multipoles using the following expression for one of the incident plane wave polarizations (with the E-field parallel to the central microstrip):

$$r = \frac{ik}{2E_0 S_L \varepsilon_0}\left[p_x - \frac{m_y}{c} + t_x + \frac{ik\left(Q^{(e)}_{xz}+T^{(Qe)}_{xz}\right)}{6} - \frac{ikQ^{(m)}_{yz}}{6c} - \frac{k^2 O^{(e)}_{xzz}}{6}\right], \quad (S1)$$

$$t = 1 + \frac{ik}{2E_0 S_L \varepsilon_0}\left[p_x + \frac{m_y}{c} + t_x - \frac{ik\left(Q^{(e)}_{xz}+T^{(Qe)}_{xz}\right)}{6} - \frac{ikQ^{(m)}_{yz}}{6c} - \frac{k^2 O^{(e)}_{xzz}}{6}\right], \quad (S2)$$

where $p$ – the electric dipole, $m$ – the magnetic dipole, $t$ – the toroidal electric dipole, $Q^{(e)}$ – the electric quadrupole (including its toroidal component), $Q^{(m)}$ – the magnetic quadrupole, $O^{(e)}$ – the electric octupole, $k$ is the wave number, $c$ is the speed of light, $E_0$ is the electric field of the normally incident plane wave at the geometric center of the meta-atom, $S_L$ is the area of the lattice unit cell ($S_L = p^2$ for the square lattice, where $p$ is the period), and $\varepsilon_0$ is the vacuum permittivity. The reflection and transmission coefficients are $R = |r|^2$ and $T = |t|^2$.

We compare the contribution of the dipoles, quadrupoles and electric octupole by plotting the reflection and transmission coefficients calculated from the corresponding multipole moments shown in Figure S2.

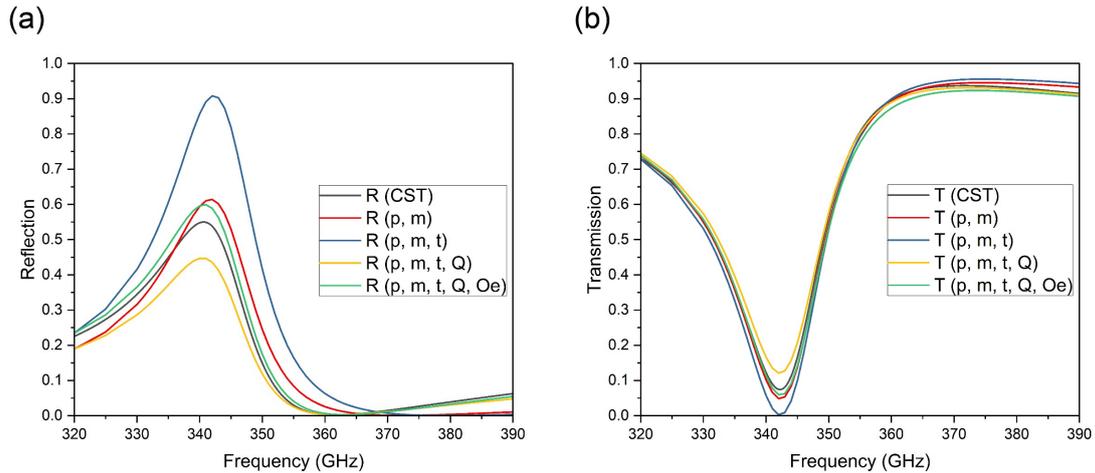

**Figure S2.** Reflection (a) and transmission (b) characteristics of the metasurface with the outer gap size (*dl*) of 160 μm calculated from exact multipoles.

By including only electric and magnetic dipoles in Equations S1 and S2, we can qualitatively reconstruct the metasurface characteristics. To further analyze the toroidal dipole contribution, we added this component to the previously considered electric and magnetic dipoles. Figure S2 illustrates that the magnitude of the resonance in reflection increases by approximately 1.5 times in this case. The electric and magnetic quadrupoles reduce the amplitude of the resonance in reflection by nearly half. Finally, the electric octupole increases the reflection coefficient to nearly the same value as when only considering the electric and magnetic dipoles. Figure S2 also shows the contributions of the above-mentioned multipoles to the transmission characteristics of the developed metasurface. These results offer important insights into the nature of the quasi-BIC resonance and demonstrate the sufficiently high accuracy of the exact expressions for multipole expansion.

Figure S3 illustrates the Q-factor and bandwidth of the quasi-BIC resonances for different outer gap sizes (*dl*) when *dc* = 40 μm and collision frequencies (gamma) corresponding to different resonator losses when *dl* = 160 μm. As the resonance disappears when approaching *dl* = 50 μm, the analysis relying on the fitting of the transmission spectrum with a Fano formula becomes inaccurate. Therefore, we considered the interval of *dl* from 90 to 180 μm. The Q-factor increases and the bandwidth decreases almost linearly with decreasing *dl*. Since we used the Drude model for gold, we also showed the Q-factor and resonance bandwidth for different collision frequencies to analyze their dependence on the resonator losses. As it can be expected, the Q-factor increases and the bandwidth decreases with decreasing collision frequency, emphasizing the need for optimization of the meta-atom and using metals with lower losses to achieve high Q-values of quasi-BIC resonances.

Figure S4 illustrates the current density within symmetric and asymmetric meta-atoms with integrated $VO_2$ patches, as shown in Figure 7, for different conductivity values of 1000 S/m (a, b) and 100000 S/m (c, d), corresponding to temperatures of 70ºC and 80ºC, respectively.

We observe an increase in current penetration into the VO$_2$ patches with increasing conductivity, coinciding with the disappearance of the quasi-BIC resonance, which requires a central gap where the electric field can be concentrated in the absence of a conductor.

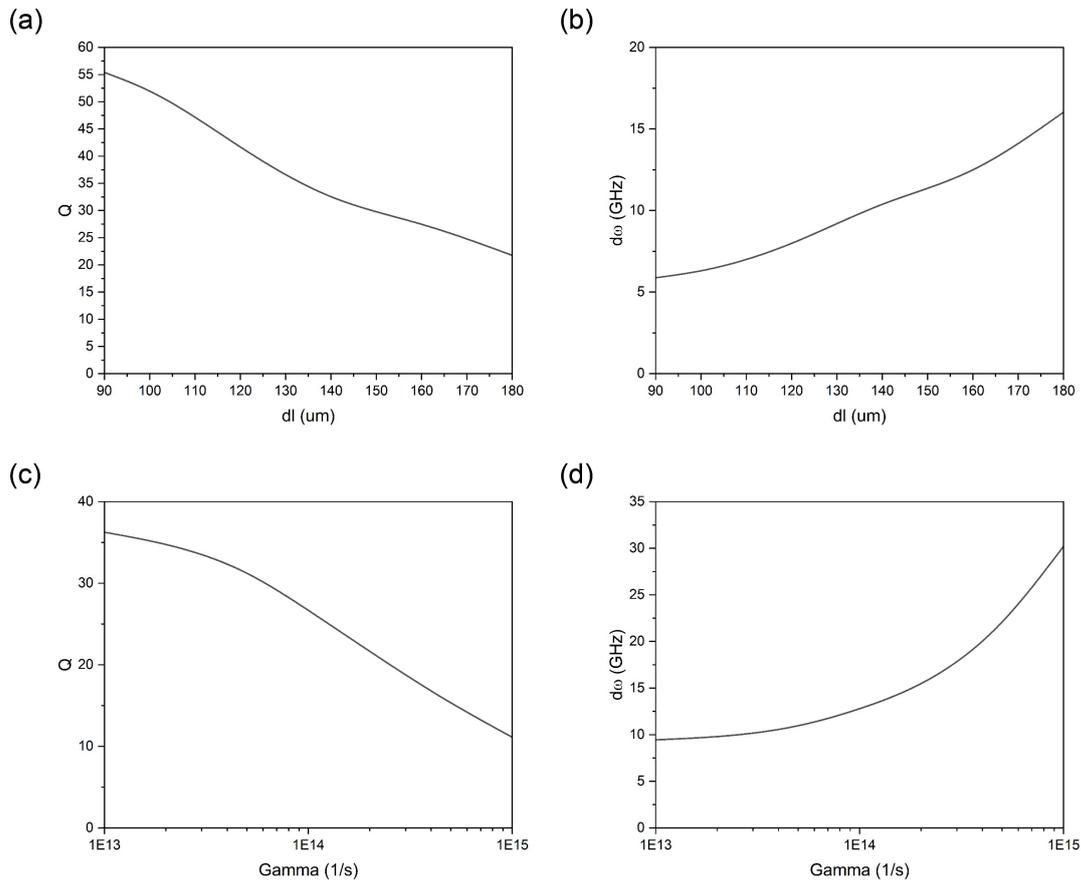

**Figure S3.** The quasi-BIC resonance Q-factor and bandwidth for different outer gap sizes (*dl*) when *dc* = 40 μm (a, b) and collision frequencies (gamma) corresponding to different resonator losses when *dl* = 160 um (c, d).

In Figure S5a and S5b, we present another design of tunable symmetric and asymmetric meta-atoms, incorporating thin film patches of vanadium dioxide in the central gap, as modeled in CST Studio. The specified parameters include the outer gap size (*dl*) of 160 μm, the central gap size (*dc*) of 40 μm, and the absorber size of 30 μm x 30 μm. The remaining meta-atom parameters remain consistent with those outlined in the article. Such design reduces the transmission change at a given frequency but allows to read the state of the absorbers without their galvanic connection to the resonators.

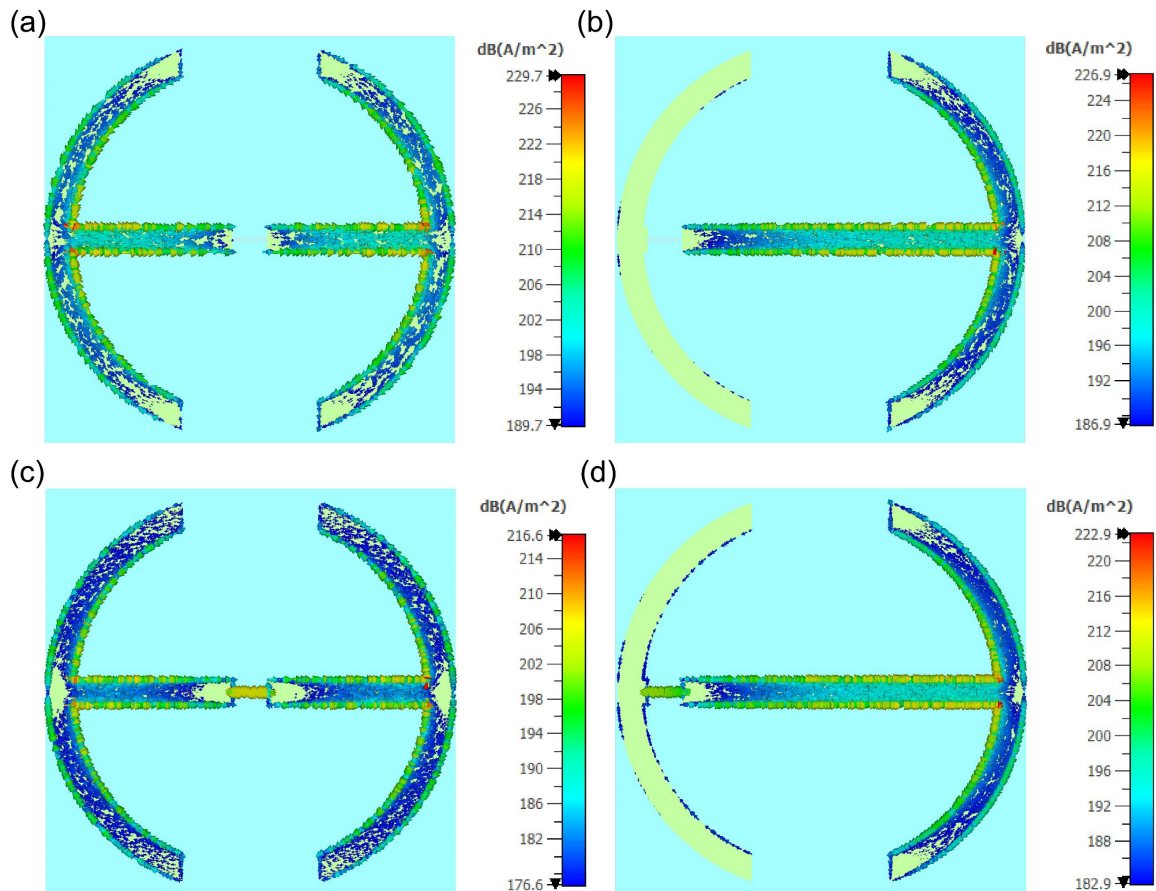

**Figure S4.** Current density within symmetric and asymmetric meta-atoms with integrated VO$_2$ patches for their different conductivity values – 1000 S/m (a, b) and 100000 S/m (c, d) corresponding to 70ºC and 80ºC, respectively.

Figures S5 (c-f) display the transmission characteristics and phase of transmitted waves for both symmetric and asymmetric meta-atoms for different values of VO$_2$ conductivity. Notably, the quasi-BIC resonance shifts to lower frequencies with increasing absorber conductivity, indicating material heating. This configuration not only enhances thermal isolation for the absorbers − a critical factor in bolometric detection − but also allows control over the quasi-BIC resonance frequency. This tunability also holds promising implications for other applications, such as reconfigurable filters.

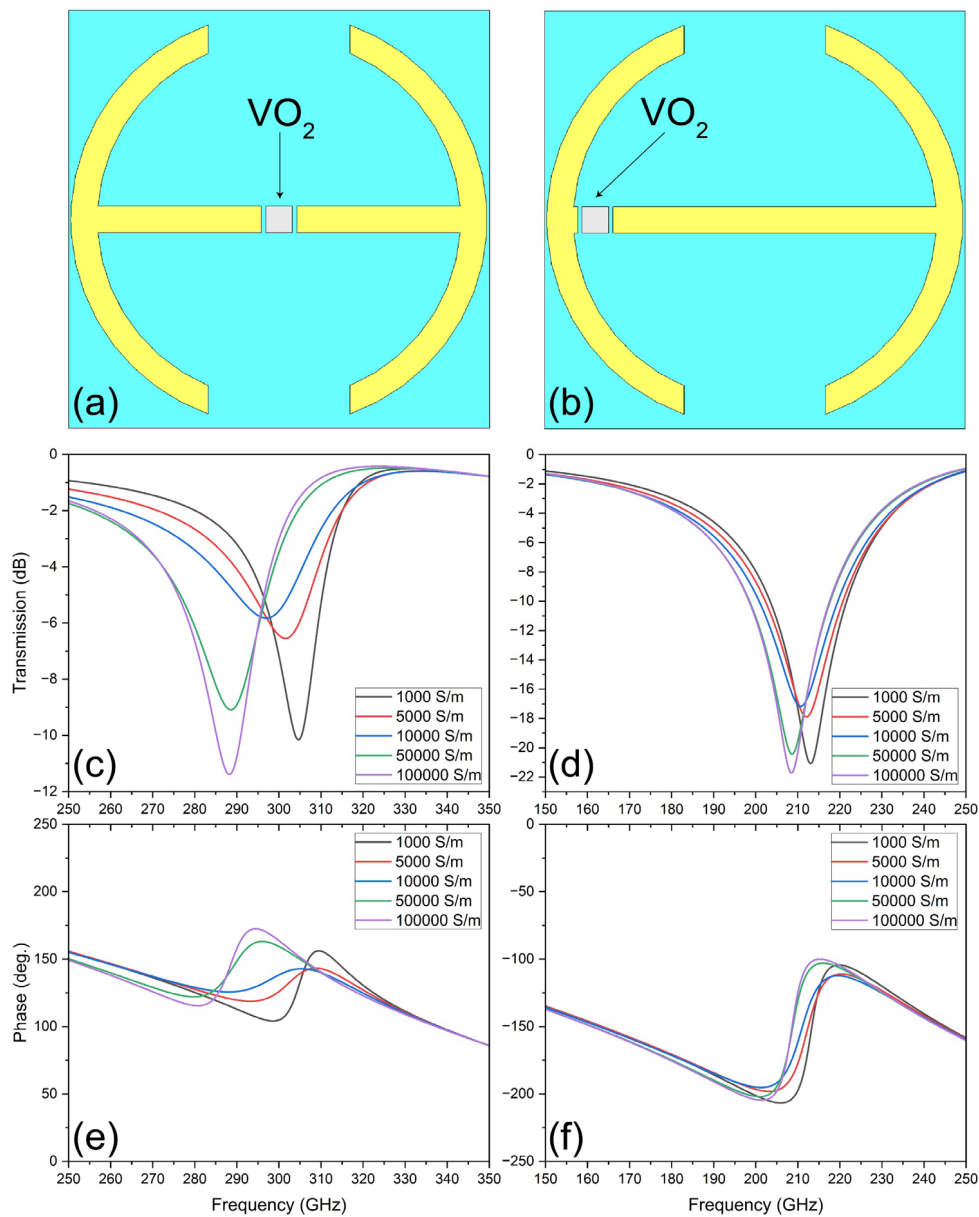

**Figure S5.** Tunable symmetric (a) and asymmetric (b) meta-atoms with the integrated thin film patches of vanadium dioxide in the central gap (without their galvanic connection to the resonators) and simulated transmission characteristics of the metasurface (c,d) and phase of transmitted waves (e,f) for different values of $VO_2$ conductivity.